# Microscopic mechanism of van der Waals heteroepitaxy in the formation of MoS$_2$/hBN vertical heterostructures

Mitsuhiro Okada[1], Mina Maruyama[2], Susumu Okada[2], Jamie H. Warner[3, 4], Yusuke Kureishi[5], Yosuke Uchiyama[5], Takashi Taniguchi[6], Kenji Watanabe[7], Tetsuo Shimizu[1], Toshitaka Kubo[1], Hisanori Shinohara[5] and Ryo Kitaura[5, *]

[1]Nanomaterials Research Institute, National Institute of Advanced Industrial Science and Technology (AIST), Tsukuba, Ibaraki 305-8565, Japan

[3]Graduate School of Pure and Applied Sciences, University of Tsukuba, Tsukuba, Ibaraki 305-8571, Japan

[3]Walker Department of Mechanical Engineering, The University of Texas at Austin, 204 East Dean Keeton Street, Austin, Texas, 78712, United States

[4]Materials Graduate Program, Texas Materials Institute, The University of Texas at Austin, 204 East Dean Keeton Street, Austin, Texas, 78712, United States

[5]Department of Chemistry, Nagoya University, Nagoya, Aichi 464-8602 Japan

[6]International Center for Materials Nanoarchitectonics, National Institute for Materials Science, Tsukuba, Ibaraki 305-0044, Japan

[7]Research Center for Functional Materials, National Institute for Materials Science, Tsukuba, Ibaraki 305-0044, Japan

Corresponding Author: R. Kitaura, r.kitaura@nagoya-u.jp

Abstract

Recent works have revealed that van der Waals (vdW) heteroepitaxial growth of 2D materials on crystalline substrates, such as hexagonal boron nitride (hBN), leads to formation of self-aligned grains, which results in defect-free stitching between the grains. However, how the weak vdW interaction causes strong limitation on crystal orientation of grains is still not understood yet. In this work, we have focused on investigation of microscopic mechanism of self-alignment of MoS$_2$ grains in vdW epitaxial growth on hBN. Through calculation based on density functional theory and the Lennard-Jones potential, we found that interlayer energy between MoS$_2$ and hBN strongly depends both on size and crystal orientation of MoS$_2$. We also found that, when size of MoS$_2$ is *ca.* 40 nm, rotational energy barrier can exceed ~ 1 eV, which should suppress rotation to align crystal orientation of MoS$_2$ even at growth temperature.

Introduction

Two-dimensional (2D) materials have been attracting much attention due to their fascinating properties and possible applications for nanoelectronics and photonics.[1-2] 2D group-VI transition metal dichalcogenides (TMDs: $MoS_2$, $WS_2$, $MoSe_2$, $WSe_2$, *etc.*), in particular, are outstanding because they can have sizable bandgap (1.5-2.0 eV) that is absent in graphene.[3-4] In addition, TMDs with 2H phase, at the monolayer limit, possess direct gap,[4-5] where spin and valley degrees of freedom are coupled due to inversion symmetry breaking.[6] These properties make TMDs as excellent platforms not only for exploring fundamental physics at the 2D limit but also for the next generation optoelectronic devices.[7]

To explore the fascinating possibilities of TMDs, wafer-size single-crystalline TMD films are required. Electronic and optoelectronic applications needs arrays of large number of devices, and small flakes prepared with the mechanical exfoliation are not compatible with these applications; typical size of flakes obtained with the mechanical exfoliation is several tens of micrometers.[2-3] Also, grain boundaries (GBs) in polycrystalline films can strongly degrade electrical and optical properties through introducing defect-induced mid-gap states at the GBs.[8-9] The mid-gap states cause extrinsic carrier scattering and non-radiative decay of excitons, which leads to degradation of carrier mobility and quantum yield of TMD films.[9-10] Single-crystalline large-area TMD films are, therefore, important to realize high-performance next generation electronic and optoelectronic devices with TMDs.

One of the strategies to achieve a large-area single-crystalline TMD film is to align domains during a crystal growth process. Crystal growth techniques, such as chemical vapor deposition (CVD), has been widely used to obtain large-area monolayer TMD films that are very difficult to obtain by the mechanical exfoliation method.[9, 11-12] CVD-grown large-area TMD films are, however, polycrystalline films with large amount of grain boundaries (GBs) caused by the merging of misaligned grains during a growth process.[13-14] This is because polynuclear growth, where nucleuses with various orientations simultaneously form, inevitably occurs in the CVD growth of TMDs.[9] To suppress this unwanted formation of GBs in the CVD growth, one of the possible approaches is utilization of seamless merging of aligned grains.

A promising approach for the alignment of TMD domains during CVD processes is to use substrate-TMD interaction. In the past few years, CVD growth of monolayer TMDs on various crystalline substrates, such as Au(111), $Al_2O_3$, graphite and hexagonal boron nitride (hBN), have been reported.[15-22] Although there is a large lattice mismatch between TMDs and a substrate (*e.g.* ~26% in the case of $MoS_2$ and hBN), the substrate-TMD interaction precisely aligns TMD domains, limiting the stacking angles between two domains to 0 and 60º.[16-18] Recent works have revealed that boundaries between two aligned $MoS_2$ domain show perfect defect-free stitching to form a single-crystalline domain.[23] The substrate-induced alignment is, therefore, crucial for realization of large-area monolayer TMD films for TMD-based electronic and optoelectronic devices.

The mechanism of substrate-induced alignment is still unclear although experimental works have clearly shown the alignment of TMD domains on various crystalline substrates. The alignment in crystal orientation is caused by van der Waals (vdW) interaction between TMDs and crystalline substrates, and this is essentially different from epitaxial growth of compound semiconductor heterostructures, where direct formation of chemical bond plays an important role.[24] Previous papers have revealed that substrate-TMD interaction energy shows stacking angle dependence, which can lead to aligned TMD crystals.[17, 20] It is still, however, unknown that how non-bonding weak interaction, such as vdW force, cause strong limitation on relative orientation during high-temperature CVD growth process, where small nuclei gradually grow to form aligned large domains.

In this work, we focus on the microscopic mechanism of the substrate-induced alignment in CVD growth of monolayer $MoS_2$ on hBN substrates. Detailed analyses on relative orientation of $MoS_2$ crystals and hBN with electron diffraction have confirmed that relative orientations between CVD-grown $MoS_2$ and underlaying hBN substrate are limited to 0 and 60°. To elucidate the mechanism of the self-alignment, we have calculated cluster-size-dependent interlayer-energy landscapes with density functional theory (DFT) and Lennard-Jones (LJ) pair potential. In both cases, we found that the interaction energy shows significant stacking-angle and cluster-size dependence. Although the stacking angle of 0/60° is not always the most stable configuration, we found, up to size of $Mo_{972}S_{1944}$, that the most stable configurations with the stacking angle of 0/60° appear periodically, whose period is close to the moiré period between $MoS_2$ and hBN. We also found that the rotational energy barrier (the difference between interaction energy at 0° and at the angle corresponding to the nearest local maximum to 0°) is comparable to thermal energy at growth temperature when the cluster size reaches ~ 40 nm. These indicates that small clusters, at the early stage of growth, rotate to change stacking angle and then the rotational degree of freedom is gradually lost as cluster size increase. This finding provides a basic to realize defect-free, large-scale 2D films onto these crystalline substrates.

Results and discussion

In order to grow $MoS_2$ onto hBN, we used the multi-furnace CVD method with elemental sulfur and molybdenum trioxide as precursors. hBN substrate is obtained by the mechanical exfoliation method with a bulk hBN crystal grown by high-pressure, high-temperature method.[25] Details on the growth procedure are shown in method section and a previous report.[26] Figure 1a shows a typical optical image of $MoS_2$ grown on hBN ($MoS_2$/hBN), where the hexagonal shaped contrasts correspond to single-layer $MoS_2$.[26-27] The grown $MoS_2$ grains with typical size of ~ 5 μm are placed randomly on the hBN substrate; a schematic image of MoS$_2$/hBN is shown in Fig. 1b. To confirm the layer number of MoS$_2$, we measured Raman spectra at room temperature with excitation energy of 2.33 eV. Figure 1c shows a Raman spectrum, where two characteristics peaks (383.7 and 404.7 cm$^{-1}$) originating from in-plane (E´ mode) and out-of-plane vibrational mode (A´$_1$ mode) exist.[28] The frequency difference between these two peaks is 21.1 cm$^{-1}$, which is slightly larger than the reported value of monolayer MoS$_2$.[29] Figure 1d shows a typical photoluminescence (PL) spectrum of the MoS$_2$/hBN at room temperature. As seen in Fig. 1d, single-peak PL emission centered at 1.884 eV is observed, which is consistent to those in previous reports.[30-31] The full-width half-maximum (FWHM) of the obtained PL peak is about 43 meV, which is significantly smaller than that of monolayer MoS$_2$ grown onto SiO$_2$/Si substrate (55 meV),[31] indicating inhomogeneous broadening arising from substrates is suppressed in MoS$_2$/hBN.[26] Atomic force microscope observations shown in Fig. S1 is also consistent to monolayer MoS$_2$.[17]

As shown in Fig. 1a, we can see that all MoS$_2$ domains have only two orientations, where 60° rotation of one orientation matches the other orientation. The observed limitation in orientation of MoS$_2$ results from the existence of strict relationship in crystallographic orientation between MoS$_2$ and hBN. In fact, a typical selected area electron diffraction (SAED) pattern shown in Fig. 1d clearly shows two sets of six-fold-symmetric spots with almost same orientation; the larger (smaller) six-fold-symmetric pattern originates from MoS$_2$ (hBN).[15-16] This means that relative crystallographic angle between MoS$_2$ and hBN is limited to two orientations, 0 and 60°, and statistics on frequency of the two stacking angles are summarized in Table 1. As clearly seen, ratio between two stacking angles is nearly 1:1, which is consistent to previous studies.[16] The slight deviation from 1:1 probably originate from statistical error or defect-controlled nucleation of MoS$_2$ on hBN.[32]

To investigate the mechanism of vdW heteroepitaxial growth of MoS$_2$/hBN, we calculated cluster-size-dependent interaction-energy landscapes by DFT. We calculated four MoS$_2$/hBN with different cluster size: the corresponding models are shown in Fig. 2a. Detailed calculation method is shown in the Methods section. Figure 2b shows the total energy of MoS$_2$ on hBN as a function of the stacking angle. As clearly seen, the energy is sensitive not only to the cluster size but also to the stacking angle. For the smallest MoS cluster, the cluster prefers the stacking angle of approximately 15°, reflecting the local atomic arrangement between MoS and hBN. The increase of the cluster size causes additional energy minima, owing to the increase of the preferential interlayer atomic arrangements arising from the lattice mismatch. Indeed, for the largest cluster, we found two global minima at the angles of 10 and 60° in addition to the two local minima, still reflecting the local atomic arrangement between them. Thus, with further increasing the flake size, MoS cluster possess some particular orientations by the averaged vdW interaction.

To have deeper insight on mechanism of the vdW-mediated epitaxial growth of $MoS_2$ onto hBN, we extended calculation of interaction-energy landscapes to larger clusters with LJ pair potential. A simple pair potential, such as LJ potential, allows us to calculate interaction energy between a cluster of $MoS_2$ and hBN, which are composed of large number of atoms. In this work, we calculated large $MoS_2$/hBN stacked structures up to $Mo_{972}S_{1944}$/$B_{5400}N_{5400}$, which is beyond the scope of DFT calculations. In LJ pair potential, interaction energy between a TMD cluster and a hBN substrate, $E^{inter}$, are described as a summation of the LJ potential.

$$E^{inter} = \sum_{i=1}^{N_{TMD}} \sum_{j=1}^{N_{hBN}} \left\{ 4\epsilon_{ij} \left[ \left(\frac{\sigma_{ij}}{r_{ij}}\right)^{12} - \left(\frac{\sigma_{ij}}{r_{ij}}\right)^{6} \right] \right\} \quad (1)$$

$N_{TMD}$, $N_{hBN}$ are number of the atoms in the TMDs and hBN, respectively. $\varepsilon_{ij}$ and $\sigma_{ij}$ are the LJ parameters corresponding to interactions between the $i$th and $j$th atom in the system, and $r_{ij}$ is the distance between the $i$th and $j$th atom in the system. We calculated stacking angle-dependent energy landscapes with various different sizes of $MoS_2$. Note that $MoS_2$ and hBN are treated as rigid bodies and intralayer energy, such as bending chemical bonds, is ignored. The LJ parameters in equation (1) were estimated with Lorentz-Berthelot combining rules, where $\sigma$ and $\varepsilon$ between different kind of atoms, $\sigma_{AB}$ and $\varepsilon_{AB}$, were calculated with those between same kind of atoms, $\sigma_{AA}$, $\sigma_{BB}$, $\varepsilon_{AA}$, and $\varepsilon_{BB}$, as follows.

$$\sigma_{AB} = \frac{\sigma_{AA} + \sigma_{BB}}{2} \quad (2a)$$

$$\epsilon_{AB} = \sqrt{\epsilon_{AA}\epsilon_{BB}} \quad (2b)$$

The LJ parameters of Mo-Mo, S-S, B-B, and N-N ($\sigma_{MoMo}$, $\sigma_{SS}$, $\sigma_{BB}$, $\sigma_{NN}$, $\varepsilon_{MoMo}$, $\varepsilon_{SS}$, $\varepsilon_{BB}$, an $\varepsilon_{NN}$), which is needed to calculate $\sigma_{MoB}$, $\sigma_{MoN}$, $\sigma_{SB}$, $\sigma_{SN}$, $\varepsilon_{MoB}$, $\varepsilon_{MoN}$, $\varepsilon_{SB}$, and $\varepsilon_{SN}$ are taken from references;[33-34] Table 2 shows the obtained LJ parameters.

To confirm validity of the LJ parameters obtained, we have checked consistency between interaction-energy landscapes calculated with LJ pair potential and ab-initio DFT calculation. Figure S2 shows stacking angle dependences of interaction energy calculated with LJ pair potential; a large hexagonal 2-layer hBN ($B_{2700}N_{2700} \times 2$) and MoS clusters used in DFT calculations are employed in this calculation. As clearly seen, calculation based on LJ potential qualitatively reproduce the results obtained with DFT (Fig. 2b). Having reproduce the DFT results successfully, we conducted LJ-potential-based calculation to obtain interaction energy of $MoS_2$/hBN up to $Mo_{972}S_{1944}$/$B_{5400}N_{5400}$.

Figure 3a shows the scheme to calculate interaction energy landscapes of $MoS_2$/hBN. First, we put a hexagonal $Mo_{3n^2}S_{6n^2}$ ($n$ = 1, 2, …, 18) cluster on a hexagonal bilayer hBN ($B_{2700}N_{2700} \times 2$) with a certain stacking angle (Fig. 3a i)), and then, we search for the ground minimum configuration at the stacking angle by simulated annealing (SA) method (Fig. 3a ii) and iii)). The SA-based ground-minimum search have been repeated with various stacking angles to calculate an interaction-energy landscape for a certain cluster size of $MoS_2$. This process has been repeated with various size of clusters, and obtained interaction-energy landscapes for 18 different clusters of $MoS_2$ are shown in Fig. 3b; the most stable configurations found for the stacking angle of 0° are shown in Fig. S3.

As shown in the Fig. 3b, the interaction-energy landscapes show several local minimums locate at various different stacking angles, and different cluster size of $MoS_2$ gives different positions of local minimums. Notably, a structure with a stacking angle of 0 or 60° does not always correspond to the ground minimum in the corresponding the interaction-energy landscape. The cluster-size dependency of interaction energy at stacking angle of 0° oscillates periodically (Fig. 3c); clusters containing 3, 75, 243, 432, and 768 Mo atoms ($n$ = 1, 5, 9, 12, and 16, respectively) have large interaction energy compared with those of clusters with similar size. It should be noted that the periodicity of oscillations in interaction energy seen in Fig. 3c is roughly close to that of moiré superlattice in $MoS_2$/hBN with stacking angle of 0°; the periodicity of the moiré superlattice is 11.96 angstrom (the moiré lattice constant $\lambda$ and the relative rotation angle $\theta$ of the moiré pattern with respect to the hBN lattice can be calculated with the following equation:[35]

$$\lambda = \frac{(1+\delta)a}{\sqrt{2(1+\delta)(1-cos\phi)+\delta^2}} \qquad (3)$$

$$\tan\theta = \frac{sin\phi}{(1+\delta)-cos\phi} \qquad (4)$$

where $\delta$, $a$, and $\phi$ represent the lattice mismatch between $MoS_2$ and hBN, the hBN lattice constant, and the relative rotation angle, respectively). This can be understood through visualizing contribution of each atom in $MoS_2$ clusters to $E^{inter}$. Figure 3e shows contribution of each $S_2$ pairs in the $Mo_{75}S_{150}$ cluster ($n$ = 5) to $E^{inter}$; the contribution is evaluated with sum of LJ potential energy between a $S_2$ pair in a cluster with total atoms in the hBN. As shown in the Fig. 3e and 3f, $S_2$ pairs that contribute to enlarge $E^{inter}$ appears with periodicity of the moiré superlattice because the stable configuration between $S_2$ and hBN, which significantly contribute to $E^{inter}$, appears with moiré superlattice period.

In the calculation of the interlayer-energy landscapes, we searched for the most stable configuration at each stacking angle. If there is large energy barrier between the most stable configuration and configurations corresponding to local minimum, it is not possible for $MoS_2$ clusters to find the most stable configuration. To see this, the interaction energy landscape on position of a cluster ($n$ = 16: $Mo_{768}S_{1536}$) were calculated (Fig. 3d). As clearly seen in the figure, potential barrier does not exist for translational motion of the $Mo_{768}S_{1536}$ cluster, and this means that the $MoS_2$ cluster can move to find the most stable configuration. This conclusion can be extended to larger $MoS_2$ clusters, which contains larger number of moiré units, because interaction energy can be roughly proportional to the number of moiré units and cluster size is not expected to alter the shape of potential profile (related discussion is given at the next paragraph). Significantly small energy barrier for translational motion in large systems, which is called superlubricity, has been observed in incommensurate systems, such as twisted graphite flakes,[36-37] and large MoS$_2$ clusters are also expected to move on a hBN substrate. These results strongly suggest that, at the early stage of the growth, small MoS$_2$ clusters formed would rotate and move to find the most stable configuration. Further discussions about the energy landscape, which includes crystal shape effect can be seen in the supporting information.

The next question is that "when does a MoS$_2$ cluster stop rotating and align?" To address this question, we focused on cluster-size dependent evolution of rotational energy barrier at fixed stacking angles; rotational energies were calculated as the difference between interaction energy at 0° and at the angle corresponding to the nearest local maximum to 0°. As shown in Fig. 3c, the stable and unstable atomic configurations appear periodically with moiré period, and hence rotational energy barriers can roughly be divided into the moiré contribution ($E^{moiré}$ × number of moiré unit cell) and the edge contribution. While the edge contribution cannot be neglected compared to the moiré contribution when cluster size is small, the moiré contribution becomes more and more dominant when cluster size becomes large; the moiré/edge contribution should depend quadratically/linearly on the cluster size. Rotational energy barriers of large MoS$_2$ clusters and hBN can, therefore, be estimated by the moiré contribution alone. Figure 4a shows a stacking angle dependence on total energy of a Mo$_{243}$S$_{486}$ cluster ($n$ = 9). As shown in the figure, there is the global minimum and a local maximum at 0 and 4.5°, respectively. Corresponding mappings of interaction energy are shown in Fig. 4b and 4c, and we extracted 10 moiré unit cells for each cluster; moiré unit cells are shown as dotted lines in Figs. 4b and c. The rotational energy barrier from the moiré contribution is calculated as ~0.1 meV/(MoS$_2$ unit cell), yielding the rotational energy barrier of a large cluster containing ~1 × 10$^4$ MoS$_2$ unit cells (cluster size of ~40 nm) as ~1 eV. Compared with thermal energy of typical CVD growth temperature (~100 meV), the rotational energy barrier is large enough, thereby suppressing rotation of clusters.

There are local or global energy minimums around 10~20 and 40~50° at the calculated potential landscape shown in Fig. 3b. These stacking angles that correspond to the interaction energy minimums, however, have not been experimentally observed as shown Fig. 1 and Table 1. This discrepancy is probably explained by variability of the stable stacking angle and translational position. As shown in Fig. 5a, the stacking angles at the local/global minimums around 10~20 and 40~50° significantly change depending on the cluster size. Furthermore, cluster size evolution also alters the stable translational position of MoS$_2$ on hBN. To see this, we have calculated stacking-angle dependent interaction energy evolution of the MoS$_2$ cluster with fixed center positions; the center of the MoS$_2$ clusters locates on B or N atoms (Fig. 5b). As clearly demonstrated, both the stable stacking angle and translational position varies depending on the cluster size, while the stacking angle of 0/60° always correspond to local or global minimum. Thus a cluster can escape the energy minimum locates at 10~20° as the cluster grows, leading to fall into the experimentally observed stacking angle of 0/60°.

In summary, we investigate the mechanism of vdW heteroepitaxy of $MoS_2$/hBN through calculation of interaction-energy landscapes. We found that interaction-energy landscapes strongly depend on size and stacking angle of clusters. We also found that the stacking angle of 0/60° always corresponds to a local or the ground minimum configuration. The energy barrier for 0/60° configuration evolves as the cluster size grows, reaching ~ 1 eV when cluster size is around 40 nm. These findings suggest that 1) at very early stage of the growth, a small $MoS_2$ cluster can rotate to find stable configurations, and 2) the rotational degree of freedom is gradually suppressed as the cluster grows, and finally stops around ~40 nm. Our results can be a basis of controlling vdW heteroepitaxy to achieve continuous, defect-free 2D film onto crystalline substrate.

Methods
CVD growth of $MoS_2$ on hBN.
We have grown monolayer $MoS_2$ onto exfoliated hBN flakes by the CVD growth method. hBN flakes were prepared by the mechanical exfoliation method on a quartz substrate. As precursors, we used molybdenum trioxide ($MoO_3$, Sigma-Aldrich, 99.5%) and elemental sulfur (Sigma-Aldrich, 99.98%). S and a quartz substrate with hBN flakes were loaded into a quartz tube with an inner diameter of 26 mm. $MoO_3$ was placed in a quartz tube with an inner diameter of 10 mm, which was placed inside the larger diameter quartz tube, to avoid unwanted reaction between S and $MoO_3$. And then, under an Ar flow of 200 sccm, we heated the quartz tubes with a three-zone electric furnace at 200, 750, and 1100 °C for S, $MoO_3$, and the substrate, respectively. A typical growth time is 20 min.

$MoS_2$ Characterization.
Optical images were taken with a typical optical microscope (Leica DM 2500 M and Nikon Eclipse ME600). Raman and PL measurements were performed by using a confocal Raman microscope (Renishaw inVia) with excitation energy of 2.33 eV. We used a transmission electron microscope (JEOL 2100) operating at 200 keV to obtain SAED pattern. A $MoS_2$/hBN was transferred onto a copper grid with standard polymer-based transfer method.

DFT Calculation.
Theoretical calculations were performed using the DFT[38-39] as implemented in the program package Simulation Tool for Atom TEchnology (STATE).[40] We used the generalized gradient approximation with the Perdew-Burke-Ernzerhof functional[41-42] to describe the exchange-correlation potential energy among interacting electrons. The weak dispersive interaction between TMD flakes and hBN was treated using the vdW-DF2 with the C09 exchange-correlation functional.[43-44]

Ultrasoft pseudopotentials generated by the Vanderbilt scheme were adopted as the interaction between electrons and nuclei.[45] Valence wavefunctions and the deficit charge density were expanded in terms of plane wave basis sets with cutoff energies of 25 and 225 Ry, respectively. The atomic structures of TMDs were fully optimized until the force acting on each atom was less than $1.33 \times 10^{-3}$ HR/au. We consider the triangular MoS flakes, $Mo_6S_{14}$, $Mo_{10}S_{24}$, $Mo_{15}S_{36}$, $Mo_{21}S_{50}$, with S edges as the structural model of MoS flakes, which are adsorbed on monolayer hBN with the lateral supercell with the sizes of 5 x 5, 7 x 7, 8 x 8, and 9 x 9, respectively, possessing the lattice parameters of $a_0$ = 1.256, 1.758, 2.009, and 2.226 nm (Fig. 2a). To exclude the unphysical dipole interaction due to the electrostatic potential difference between MoS and hBN, we imposed an open boundary condition normal to the h-BN sheet using the effective screening medium method,[46] which the sufficiently large normal lattice parameters in which the TMD/hBN is separated from adjacents by 1.3 nm vacuum spacing. Integration over the Brillouin zone was carried out using an equidistance mesh of 2 x 2 x 1 $k$ points.

Acknowledgement
This work was supported by JSPS KAKENHI Grant numbers 19K15403, JP16H06331, JP16H03825, JP16H00963, JP15K13283, JP25107002, and JST CREST Grant Number JPMJCR16F3. K.W. and T.T. acknowledge support from the Elemental Strategy Initiative conducted by the MEXT, Japan, Grant Number JPMXP0112101001, JSPS KAKENHI Grant Numbers JP20H00354 and the CREST(JPMJCR15F3), JST.

Notes
The authors declare no competing financial interest.

ORCID
Mitsuhiro Okada: 0000-0002-6123-1094

Supporting Information Available.

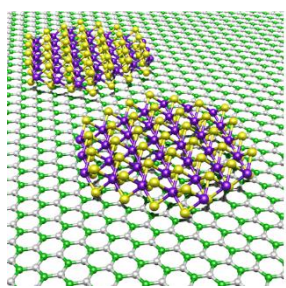 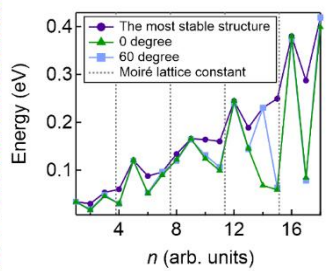

**TOC graphic**.

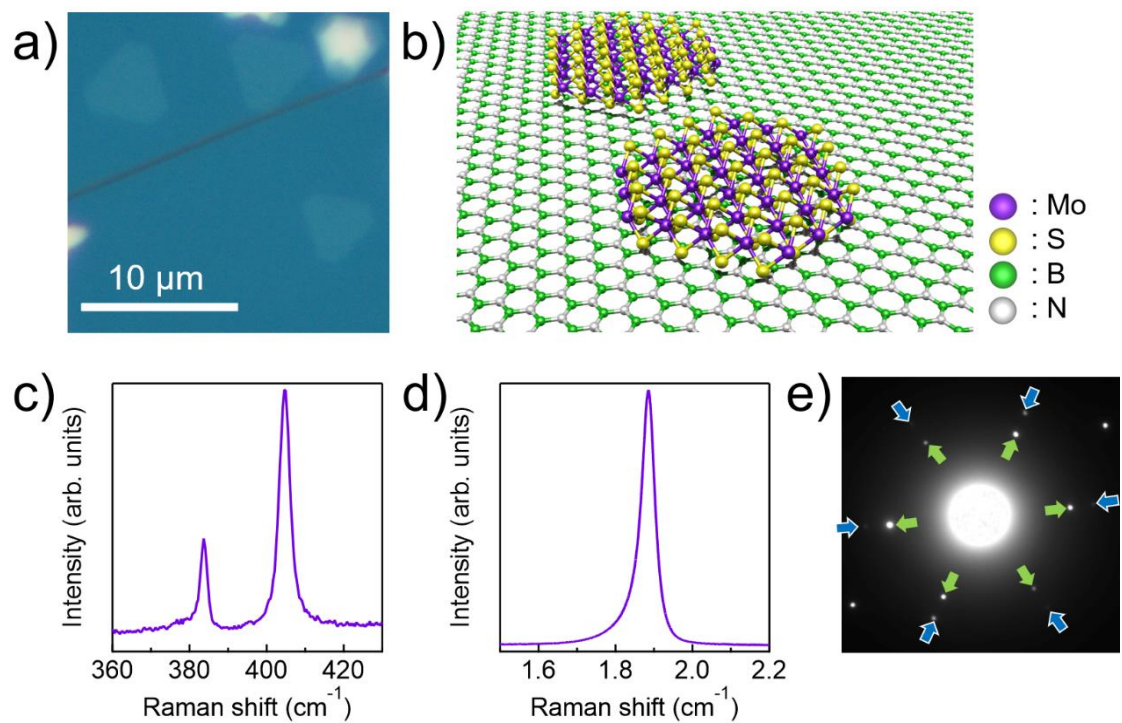

**Figure 1**. a) An optical image of CVD-grown MoS$_2$ on hBN; b) a schematic of MoS$_2$/hBN; c): a typical Raman spectrum and d): PL spectrum of the MoS$_2$ crystal shown in upper left side of Fig. 1a; e): a typical SAED pattern of a MoS$_2$/hBN. Green and blue arrows indicate diffraction spots from the MoS$_2$ and the hBN in the MoS$_2$/hBN, respectively.

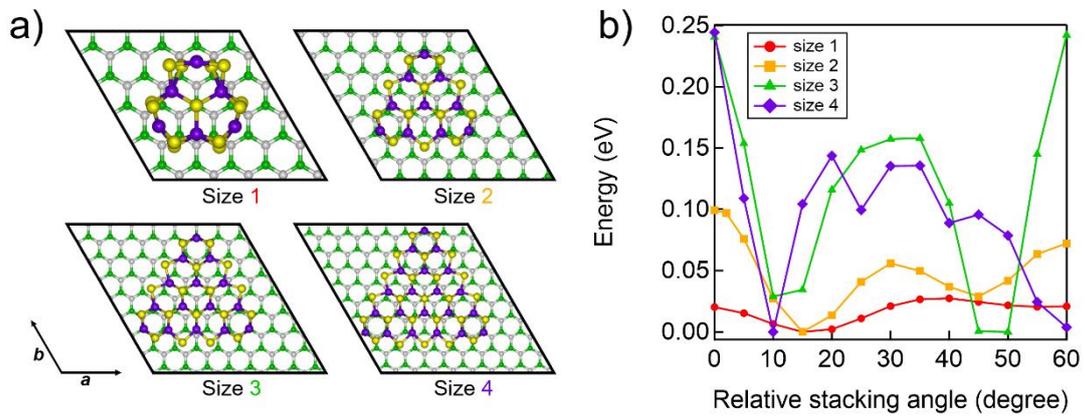

**Figure 2**. a): Schematics of the structure model used in DFT calculations. Coloring of elements are the same as that used in Fig. 1b. b): Stacking-angle-dependent total energy calculated with different cluster sizes. The total energy at the most stable stacking angle is set to zero.

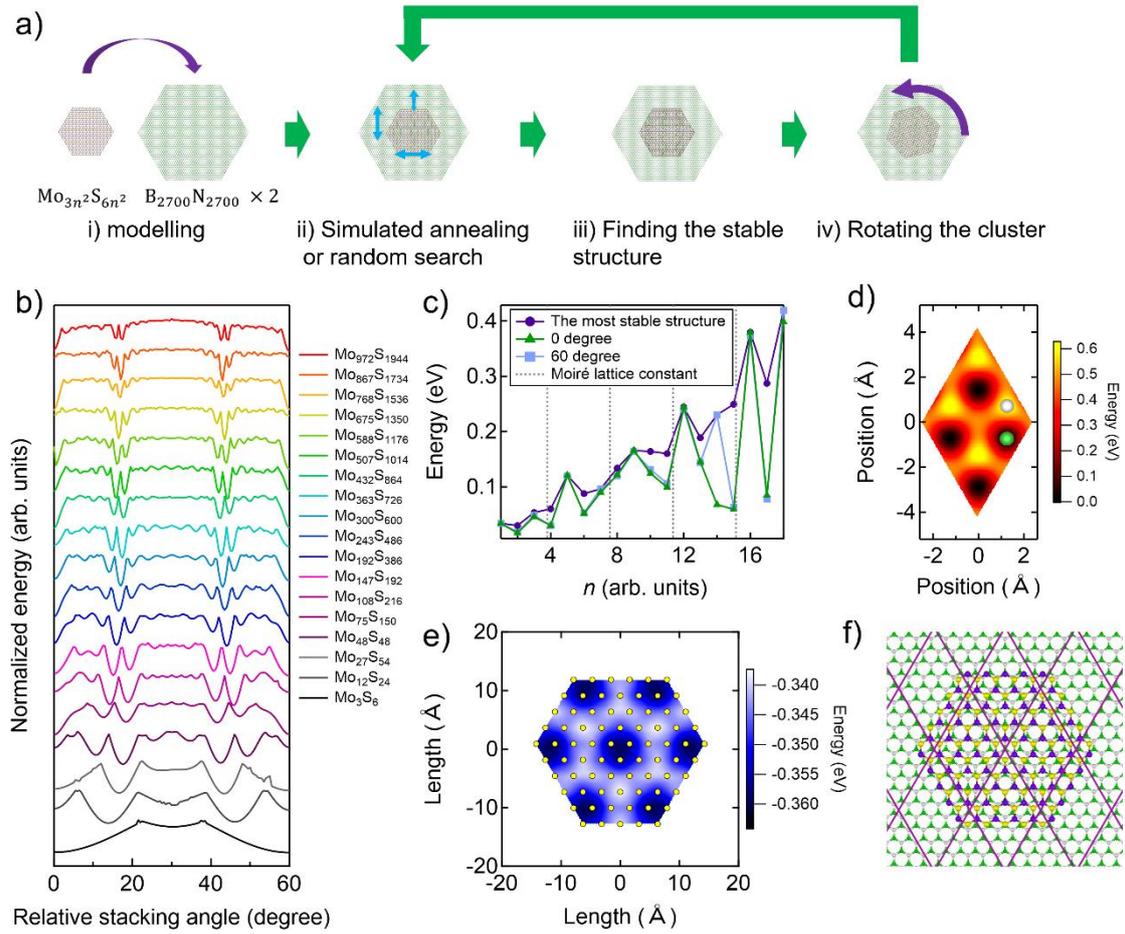

**Figure 3.** a): a schematic of the method for calculating interaction energy; b): cluster-size and stacking-angle evolutions of interaction energy; c): cluster-size dependences of absolute values of interaction energy calculated with three different stacking angles, 0°, 60°, and the most stable stacking angle (*i.e.*, absolute values of difference between the global maximum energy and energy at each stacking angle). $n$ is the term used in "$Mo_{3n^2}S_{6n^2}$" and equal to the number of S or Mo atoms (or unit cells) places at a edge of the cluster; d): an in-plane interaction-energy map of a $Mo_{768}S_{1536}$ cluster with the stacking angle of 0°. Interaction energies were calculated at various different lateral position. Grey and green sphere correspond to a position of N and B atom, respectively; e): A map showing element-decomposed interaction energies of a $Mo_{75}S_{150}$ cluster with a stacking angle of 0°. Yellow circles correspond to positions of $S_2$ pairs; f): the corresponding structure used in the calculation of the element-decomposed interaction energy in e). Purple lines indicate the moiré superlattice period. Element coloring is the same as that of Fig. 1b.

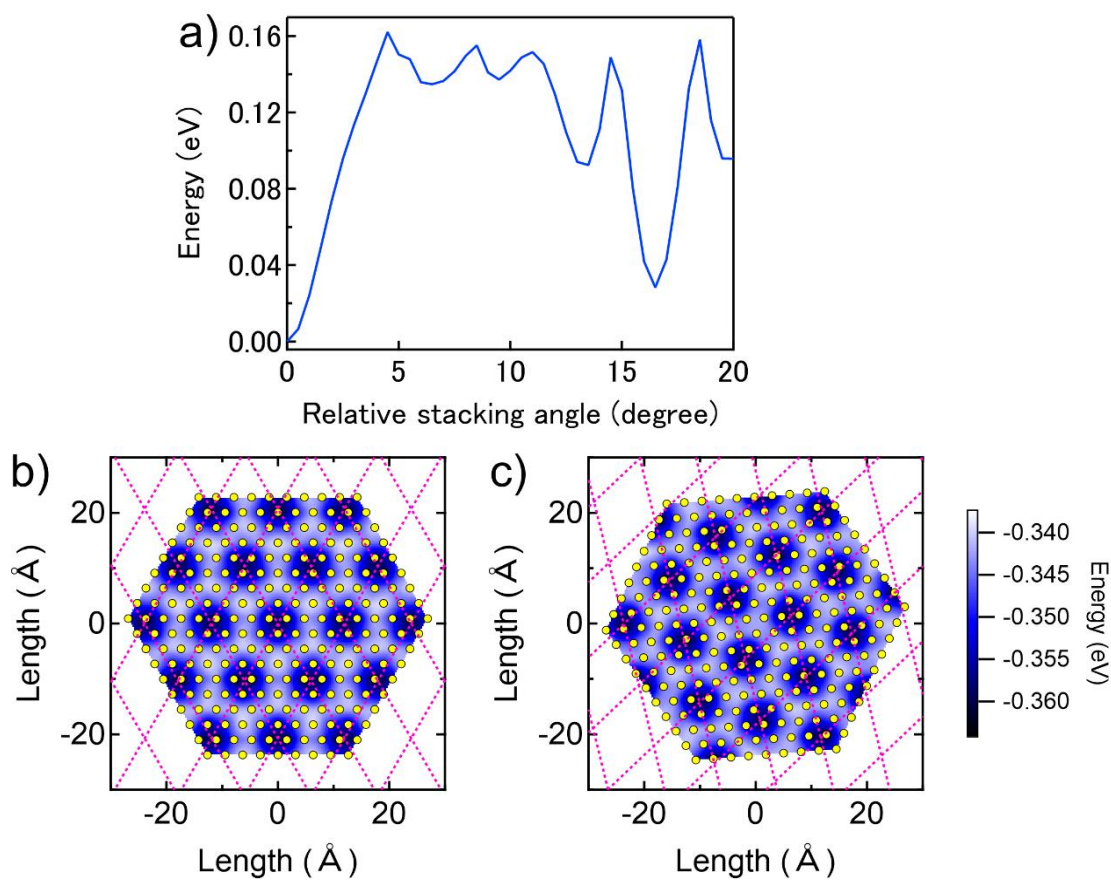

**Figure 4**. a): a stacking angle dependent interaction energy of a $Mo_{243}S_{486}$ cluster (the interaction energy at the most stable stacking angle is set to zero); b), c): maps showing element-decomposed interaction energies of $S_2$ pairs in a $Mo_{243}S_{486}$ with a stacking angle of 0º and 4.5º. Yellow circles correspond to positions of $S_2$ pairs, and magenta dotted lines correspond to the moiré superlattice period of each structure.

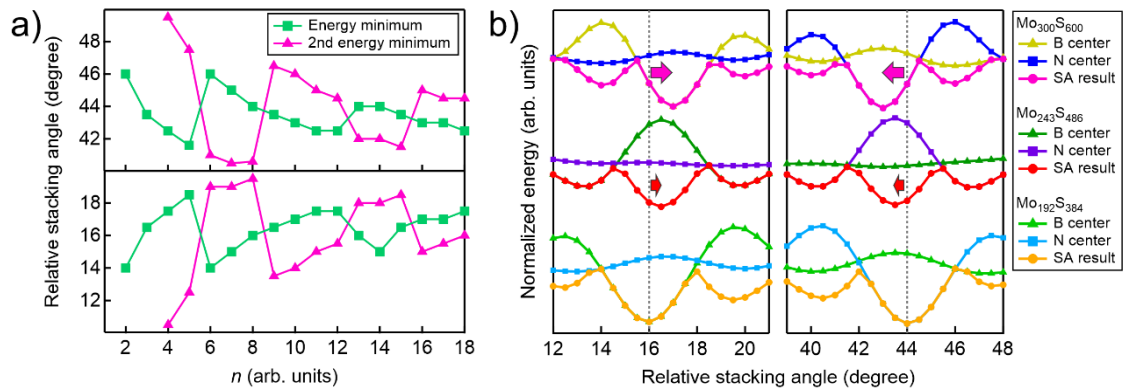

**Figure 5**. a): cluster-size-dependent stable stacking-angle evolutions around 40-50º (upper panel) and 10-20º (lower panel); b): stacking-angle and structure-dependent interaction-energy evolution of $Mo_{192}S_{384}$, $Mo_{243}S_{486}$, and $Mo_{300}S_{600}$ cluster ($n$ = 8, 9, and 10, respectively). Left and right panel show the results calculated with stacking angles of 12-21º and 39-48º, respectively. Curves labelled as "SA result" are the same as the curves shown in Fig. 3b. Black dotted lines correspond to an energy minimum position of $Mo_{192}S_{384}$.

**Table 1**. Distribution of MoS$_2$ grain orientation.

| Structure | Aligned – 0º | Aligned – 60º | Others |
|---|---|---|---|
| Number of grains | 60 | 46 | 0 |

**Table 2**. The LJ parameter.

| Interaction | ε (eV) | σ (Å) |
|---|---|---|
| Mo-B | 0.05873 | 3.002 |
| Mo-N | 0.07256 | 2.958 |
| S-B | 0.01579 | 3.411 |
| S-N | 0.01951 | 3.367 |

References

1. Novoselov, K. S.; Geim, A. K.; Morozov, S. V.; Jiang, D.; Zhang, Y.; Dubonos, S. V.; Grigorieva, I. V.; Firsov, A. A., Electric Field Effect in Atomically Thin Carbon Films. *Science* **2004,** *306* (5696), 666-669.

2. Radisavljevic, B.; Radenovic, A.; Brivio, J.; Giacometti, V.; Kis, A., Single-layer MoS2 transistors. *Nat. Nanotechnol.* **2011,** *6* (3), 147-150.

3. Mak, K. F.; Lee, C.; Hone, J.; Shan, J.; Heinz, T. F., Atomically thin MoS(2): a new direct-gap semiconductor. *Phys. Rev. Lett.* **2010,** *105* (13), 136805.

4. Kumar, A.; Ahluwalia, P. K., Electronic structure of transition metal dichalcogenides monolayers 1H-MX2 (M= Mo, W; X= S, Se, Te) from ab-initio theory: new direct band gap semiconductors. *Eur. Phys. J. B* **2012,** *85* (6), 186.

5. Kuc, A.; Zibouche, N.; Heine, T., Influence of quantum confinement on the electronic structure of the transition metal sulfide T S 2. *Phys. Rev. B* **2011,** *83* (24), 245213.

6. Xiao, D.; Liu, G.-B.; Feng, W.; Xu, X.; Yao, W., Coupled spin and valley physics in monolayers of MoS2 and other group-VI dichalcogenides. *Phys. Rev. Lett.* **2012,** *108* (19), 196802.

7. Zhang, Y. J.; Oka, T.; Suzuki, R.; Ye, J. T.; Iwasa, Y., Electrically Switchable Chiral Light-Emitting Transistor. *Science* **2014,** *344* (6185), 725-728.

8. Zhou, W.; Zou, X.; Najmaei, S.; Liu, Z.; Shi, Y.; Kong, J.; Lou, J.; Ajayan, P. M.; Yakobson, B. I.; Idrobo, J. C., Intrinsic structural defects in monolayer molybdenum disulfide. *Nano Lett.* **2013,** *13* (6), 2615-2622.

9. van der Zande, A. M.; Huang, P. Y.; Chenet, D. A.; Berkelbach, T. C.; You, Y.; Lee, G.-H.; Heinz, T. F.; Reichman, D. R.; Muller, D. A.; Hone, J. C., Grains and grain boundaries in highly crystalline monolayer molybdenum disulphide. *Nat. Mater.* **2013,** *12* (6), 554-561.

10. Ly, T. H.; Perello, D. J.; Zhao, J.; Deng, Q.; Kim, H.; Han, G. H.; Chae, S. H.; Jeong, H. Y.; Lee, Y. H., Misorientation-angle-dependent electrical transport across molybdenum disulfide grain boundaries. *Nat. Commun.* **2016,** *7* (1), 10426.

11. Lee, Y.-H.; Zhang, X.-Q.; Zhang, W.; Chang, M.-T.; Lin, C.-T.; Chang, K.-D.; Yu, Y.-C.; Wang, J. T.-W.; Chang, C.-S.; Li, L.-J.; Lin, T.-W., Synthesis of Large-Area MoS2 Atomic Layers with Chemical Vapor Deposition. *Adv. Mater.* **2012,** *24* (17), 2320-2325.

12. Lin, Z.; McCreary, A.; Briggs, N.; Subramanian, S.; Zhang, K.; Sun, Y.; Li, X.; Borys, N. J.; Yuan, H.; Fullerton-Shirey, S. K.; Chernikov, A.; Zhao, H.; McDonnell, S.; Lindenberg, A. M.; Xiao, K.; LeRoy, B. J.; Drndić, M.; Hwang, J. C. M.; Park, J.; Chhowalla, M.; Schaak, R. E.; Javey, A.; Hersam, M. C.; Robinson, J.; Terrones, M., 2D materials advances: from large scale synthesis and controlled heterostructures to improved characterization techniques, defects and applications. *2D Mater.* **2016,** *3* (4), 042001.

13. Kang, K.; Xie, S.; Huang, L.; Han, Y.; Huang, P. Y.; Mak, K. F.; Kim, C.-J.; Muller, D.;

Park, J., High-mobility three-atom-thick semiconducting films with wafer-scale homogeneity. *Nature* **2015,** *520*, 656.

14. Najmaei, S.; Liu, Z.; Zhou, W.; Zou, X.; Shi, G.; Lei, S.; Yakobson, B. I.; Idrobo, J.-C.; Ajayan, P. M.; Lou, J., Vapour phase growth and grain boundary structure of molybdenum disulphide atomic layers. *Nat. Mater.* **2013,** *12* (8), 754-759.

15. Yan, A.; Velasco, J.; Kahn, S.; Watanabe, K.; Taniguchi, T.; Wang, F.; Crommie, M. F.; Zettl, A., Direct Growth of Single- and Few-Layer MoS2 on h-BN with Preferred Relative Rotation Angles. *Nano Lett.* **2015,** *15* (10), 6324-6331.

16. Yu, H.; Yang, Z.; Du, L.; Zhang, J.; Shi, J.; Chen, W.; Chen, P.; Liao, M.; Zhao, J.; Meng, J.; Wang, G.; Zhu, J.; Yang, R.; Shi, D.; Gu, L.; Zhang, G., Precisely Aligned Monolayer MoS2 Epitaxially Grown on h-BN basal Plane. *Small* **2017,** *13* (7), 1603005.

17. Dumcenco, D.; Ovchinnikov, D.; Marinov, K.; Lazić, P.; Gibertini, M.; Marzari, N.; Sanchez, O. L.; Kung, Y.-C.; Krasnozhon, D.; Chen, M.-W.; Bertolazzi, S.; Gillet, P.; Fontcuberta i Morral, A.; Radenovic, A.; Kis, A., Large-Area Epitaxial Monolayer MoS2. *ACS Nano* **2015,** *9* (4), 4611-4620.

18. Okada, M.; Sawazaki, T.; Watanabe, K.; Taniguch, T.; Hibino, H.; Shinohara, H.; Kitaura, R., Direct chemical vapor deposition growth of WS2 atomic layers on hexagonal boron nitride. *ACS Nano* **2014,** *8* (8), 8273-8277.

19. Kobayashi, Y.; Sasaki, S.; Mori, S.; Hibino, H.; Liu, Z.; Watanabe, K.; Taniguchi, T.; Suenaga, K.; Maniwa, Y.; Miyata, Y., Growth and optical properties of high-quality monolayer WS2 on graphite. *ACS Nano* **2015,** *9* (4), 4056-4063.

20. Aljarb, A.; Cao, Z.; Tang, H.-L.; Huang, J.-K.; Li, M.; Hu, W.; Cavallo, L.; Li, L.-J., Substrate Lattice-Guided Seed Formation Controls the Orientation of 2D Transition-Metal Dichalcogenides. *ACS Nano* **2017,** *11* (9), 9215-9222.

21. Bignardi, L.; Lizzit, D.; Bana, H.; Travaglia, E.; Lacovig, P.; Sanders, C. E.; Dendzik, M.; Michiardi, M.; Bianchi, M.; Ewert, M.; Buß, L.; Falta, J.; Flege, J. I.; Baraldi, A.; Larciprete, R.; Hofmann, P.; Lizzit, S., Growth and structure of singly oriented single-layer tungsten disulfide on Au(111). *Phys. Rev. Mater.* **2019,** *3* (1), 014003.

22. Wang, S.; Wang, X.; Warner, J. H., All Chemical Vapor Deposition Growth of MoS2: h-BN Vertical van der Waals Heterostructures. *ACS Nano* **2015,** *9* (5), 5246-5254.

23. Nakanishi, T.; Yoshida, S.; Murase, K.; Takeuchi, O.; Taniguchi, T.; Watanabe, K.; Shigekawa, H.; Kobayashi, Y.; Miyata, Y.; Shinohara, H.; Kitaura, R., The Atomic and Electronic Structure of 0° and 60° Grain Boundaries in MoS2. *Frontiers in Physics* **2019,** *7* (59).

24. Koma, A., Van der Waals epitaxy—a new epitaxial growth method for a highly lattice-mismatched system. *Thin Solid Films* **1992,** *216* (1), 72-76.

25. Taniguchi, T.; Watanabe, K., Synthesis of high-purity boron nitride single crystals under

high pressure by using Ba–BN solvent. *J. Cryst. Growth* **2007,** *303* (2), 525-529.

26. Okada, M.; Kutana, A.; Kureishi, Y.; Kobayashi, Y.; Saito, Y.; Saito, T.; Watanabe, K.; Taniguchi, T.; Gupta, S.; Miyata, Y.; Yakobson, B. I.; Shinohara, H.; Kitaura, R., Direct and Indirect Interlayer Excitons in a van der Waals Heterostructure of hBN/WS2/MoS2/hBN. *ACS Nano* **2018,** *12* (3), 2498-2505.

27. Wang, S.; Rong, Y.; Fan, Y.; Pacios, M.; Bhaskaran, H.; He, K.; Warner, J. H., Shape Evolution of Monolayer MoS2 Crystals Grown by Chemical Vapor Deposition. *Chem. Mater.* **2014,** *26* (22), 6371-6379.

28. Terrones, H.; Del Corro, E.; Feng, S.; Poumirol, J. M.; Rhodes, D.; Smirnov, D.; Pradhan, N. R.; Lin, Z.; Nguyen, M. A.; Elías, A. L.; Mallouk, T. E.; Balicas, L.; Pimenta, M. A.; Terrones, M., New first order Raman-active modes in few layered transition metal dichalcogenides. *Sci. Rep.* **2014,** *4*, 4215.

29. Wu, S.; Huang, C.; Aivazian, G.; Ross, J. S.; Cobden, D. H.; Xu, X., Vapor-solid growth of high optical quality MoS(2) monolayers with near-unity valley polarization. *ACS Nano* **2013,** *7* (3), 2768-2772.

30. Gong, Y.; Lin, Z.; Ye, G.; Shi, G.; Feng, S.; Lei, Y.; Elías, A. L.; Perea-Lopez, N.; Vajtai, R.; Terrones, H.; Liu, Z.; Terrones, M.; Ajayan, P. M., Tellurium-Assisted Low-Temperature Synthesis of MoS2 and WS2 Monolayers. *ACS Nano* **2015,** *9* (12), 11658-11666.

31. Ling, X.; Lee, Y. H.; Lin, Y.; Fang, W.; Yu, L.; Dresselhaus, M. S.; Kong, J., Role of the seeding promoter in MoS2 growth by chemical vapor deposition. *Nano Lett.* **2014,** *14* (2), 464-472.

32. Zhang, F.; Wang, Y.; Erb, C.; Wang, K.; Moradifar, P.; Crespi, V. H.; Alem, N., Full orientation control of epitaxial ${\mathrm{MoS}}_{2}$ on hBN assisted by substrate defects. *Phys. Rev. B* **2019,** *99* (15), 155430.

33. Varshney, V.; Patnaik, S. S.; Muratore, C.; Roy, A. K.; Voevodin, A. A.; Farmer, B. L., MD simulations of molybdenum disulphide (MoS2): Force-field parameterization and thermal transport behavior. *Comput. Mater. Sci.* **2010,** *48* (1), 101-108.

34. Won, C. Y.; Aluru, N. R., Water Permeation through a Subnanometer Boron Nitride Nanotube. *J. Am. Chem. Soc.* **2007,** *129* (10), 2748-2749.

35. Yankowitz, M.; Xue, J.; Cormode, D.; Sanchez-Yamagishi, J. D.; Watanabe, K.; Taniguchi, T.; Jarillo-Herrero, P.; Jacquod, P.; LeRoy, B. J., Emergence of superlattice Dirac points in graphene on hexagonal boron nitride. *Nat. Phys.* **2012,** *8*, 382.

36. Dienwiebel, M.; Verhoeven, G. S.; Pradeep, N.; Frenken, J. W. M.; Heimberg, J. A.; Zandbergen, H. W., Superlubricity of Graphite. *Phys. Rev. Lett.* **2004,** *92* (12), 126101.

37. Wang, W.; Shen, J.; He, Q. C., Microscale superlubricity of graphite under various twist angles. *Phys. Rev. B* **2019,** *99* (5), 054103.

38. Hohenberg, P.; Kohn, W., Inhomogeneous Electron Gas. *Phys. Rev.* **1964,** *136* (3B), B864-B871.

39. Kohn, W.; Sham, L. J., Self-Consistent Equations Including Exchange and Correlation Effects. *Phys. Rev.* **1965,** *140* (4A), A1133-A1138.

40. Morikawa, Y.; Iwata, K.; Terakura, K., Theoretical study of hydrogenation process of formate on clean and Zn deposited Cu(111) surfaces. *Appl. Surf. Sci.* **2001,** *169-170*, 11-15.

41. Perdew, J. P.; Burke, K.; Ernzerhof, M., Generalized gradient approximation made simple. *Phys. Rev. Lett.* **1996,** *77* (18), 3865-3868.

42. Perdew, J. P.; Burke, K.; Ernzerhof, M., Generalized Gradient Approximation Made Simple [Phys. Rev. Lett. 77, 3865 (1996)]. *Phys. Rev. Lett.* **1997,** *78* (7), 1396-1396.

43. Lee, K.; Murray, É. D.; Kong, L.; Lundqvist, B. I.; Langreth, D. C., Higher-accuracy van der Waals density functional. *Phys. Rev. B* **2010,** *82* (8), 081101.

44. Cooper, V. R., Van der Waals density functional: An appropriate exchange functional. *Phys. Rev. B* **2010,** *81* (16), 161104.

45. Vanderbilt, D., Soft self-consistent pseudopotentials in a generalized eigenvalue formalism. *Phys. Rev. B* **1990,** *41* (11), 7892-7895.

46. Otani, M.; Sugino, O., First-principles calculations of charged surfaces and interfaces: A plane-wave nonrepeated slab approach. *Phys. Rev. B* **2006,** *73* (11), 115407.

**Supporting information**

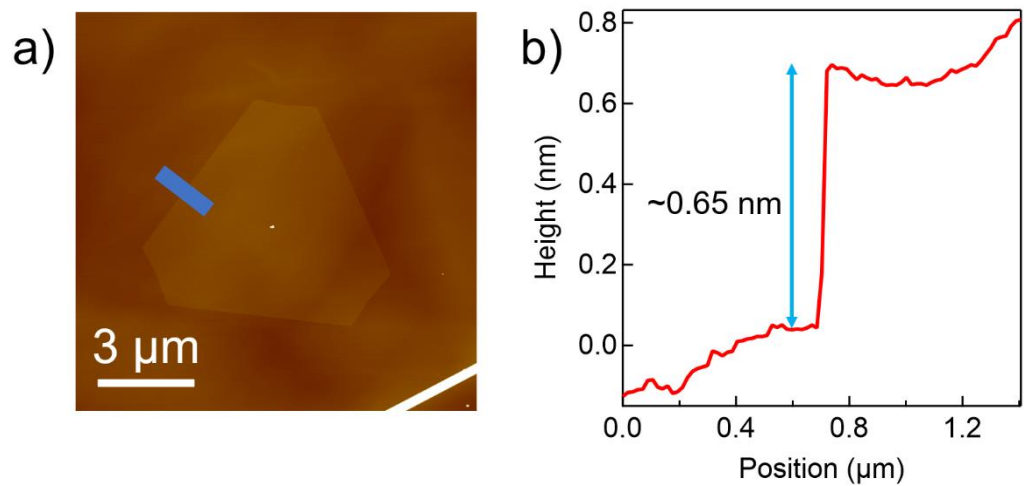

**Figure S1**. a): typical AFM image of MoS$_2$ grown on hBN; b): corresponding averaged height profile along the blue line shown in Figure S1a.

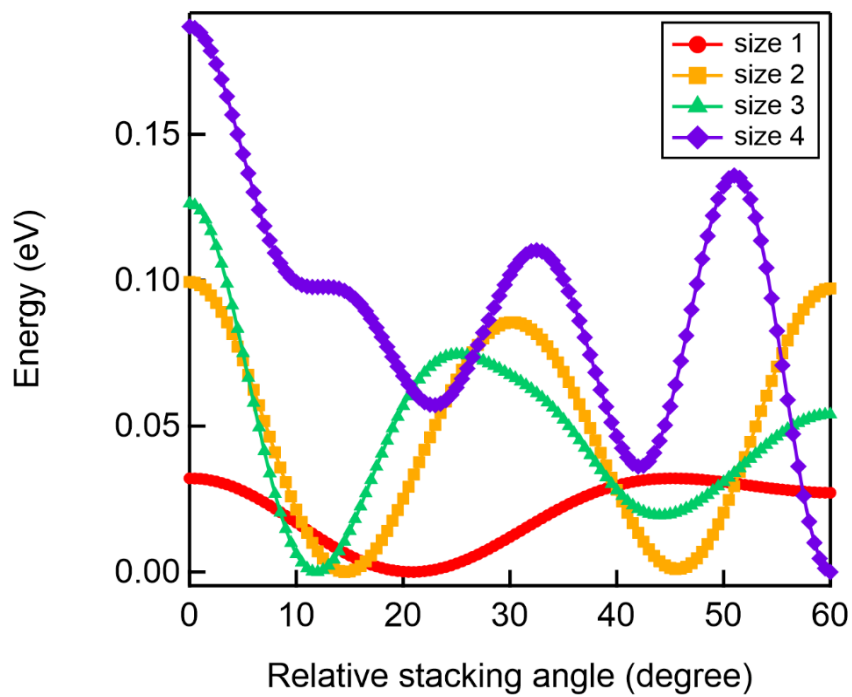

**Figure S2.** Stacking-angle dependent interaction energy of the MoS$_2$ cluster using the classical mechanics. The cluster model is same as that used in the DFT calculation (shown in Fig. 2a). The interaction energy at the most stable stacking angle is set to zero.

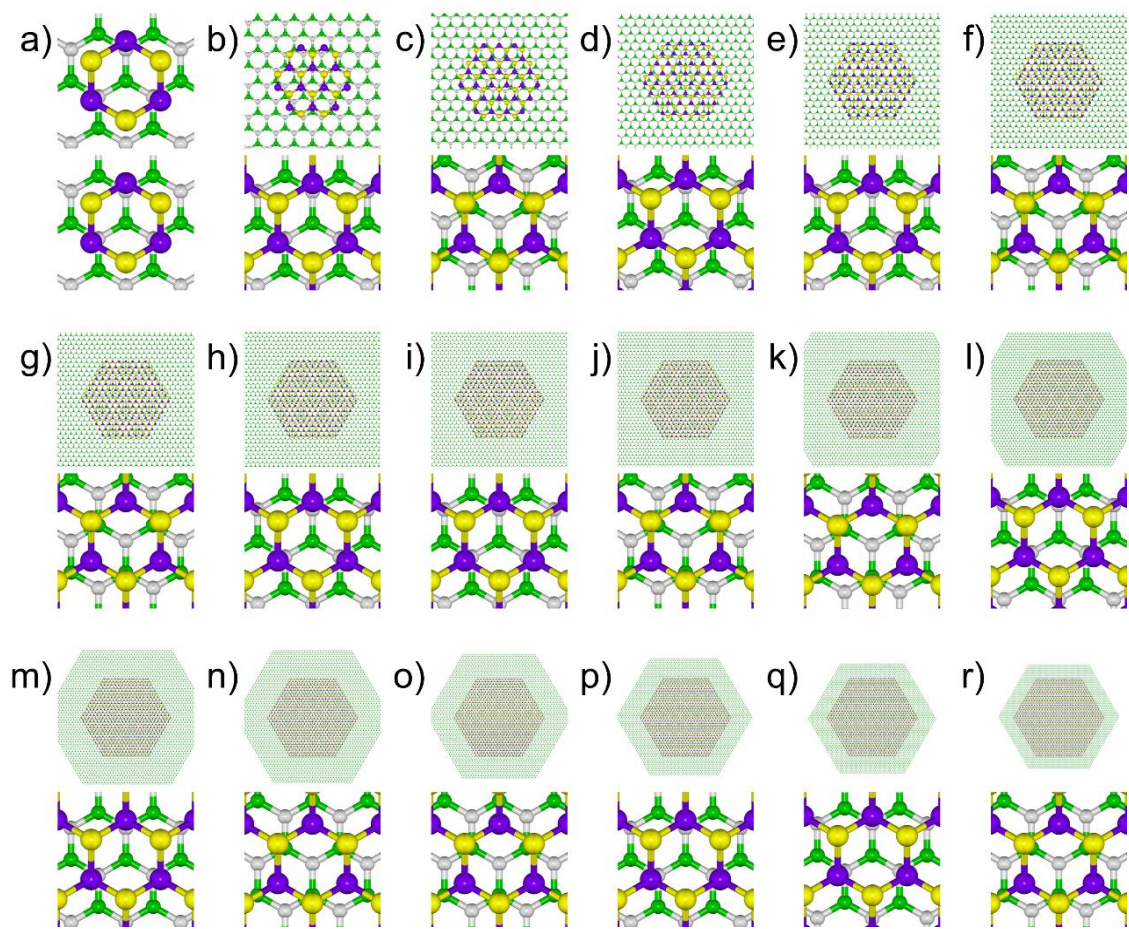

**Figure S3**. Schematic images of stable MoS$_2$/hBN structure at a stacking angle of 0°. Upper images show whole view of MoS$_2$ cluster and lower images correspond to magnified images of the center position of each clusters. A list of stable center position can be seen at Table S1. a) to r) show result of a): Mo$_3$S$_6$, b): Mo$_{12}$S2$_4$, c): Mo$_{27}$S$_{54}$, d): Mo$_{48}$S$_{96}$, e): Mo$_{75}$S$_{150}$, f): Mo$_{108}$S$_{216}$, g): Mo$_{147}$S$_{294}$, h): Mo$_{192}$S$_{384}$, i): Mo$_{243}$S$_{486}$, j): Mo$_{300}$S$_{600}$, k): Mo$_{363}$S$_{726}$, l): Mo$_{432}$S$_{864}$, m): Mo$_{508}$S$_{1016}$, n): Mo$_{588}$S$_{1176}$, o): Mo$_{675}$S$_{1350}$, p): Mo$_{768}$S$_{1536}$, q): Mo$_{867}$S$_{1734}$, and r): Mo$_{972}$S$_{1944}$ cluster. Purple, yellow, green, and grey sphere correspond to Mo, S, B, and N atom, respectively.

Table S1. Most stable center position of the MoS$_2$ cluster at a stacking angle of 0º.

| Number of Mo | Center |
|---|---|
| 3 | on B atom |
| 12 | on B atom |
| 27 | on N atom |
| 48 | on B atom |
| 75 | on B atom |
| 108 | on N atom |
| 147 | on N atom |
| 192 | on B atom |
| 243 | on B atom |
| 300 | on N atom |
| 363 | on N atom |
| 432 | on B atom |
| 507 | on B atom |
| 588 | on N atom |
| 675 | on N atom |
| 768 | on B atom |
| 867 | on B atom |
| 972 | on N atom |

**Additional discussion: effect of MoS$_2$ cluster shape**

We have checked an effect on MoS$_2$ crystal shape onto LJ potential energy. Figure S4 shows a stacking-angle dependence on total energy of triangular-shape MoS clusters (Mo$_3$S$_{14}$ and Mo$_{351}$S$_{750}$) using the same method used to obtain Figure 3b. The obtained energy evolution (Fig. S4a) is similar to the results obtained from hexagonal-shape one: 1): 0 or 60° stacking can be local or global energy minimum with crystal size evolution; and 2): there are energy minimums around 10-20 and 40-50° in the result of Mo$_{351}$S$_{750}$ cluster. Therefore, we have concluded that MoS cluster shape does not significantly affect the energy landscape obtained from the LJ potential.

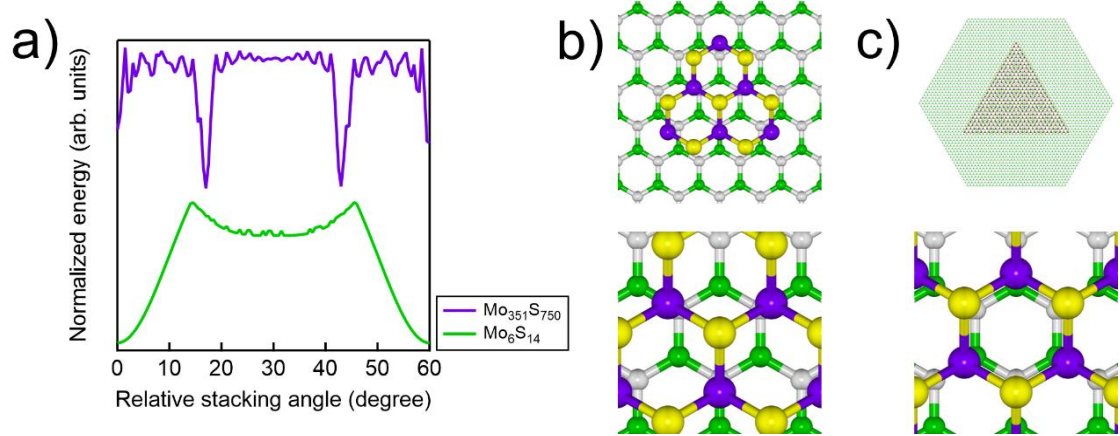

**Figure S4**. a): cluster size and relative stacking angle evolution of total energy using triangular-shape MoS cluster; b) and c): schematics of stable (b): Mo$_6$S$_{14}$ and c): Mo$_{351}$S$_{750}$)/hBN structure at 0°. Upper and lower image shows whole view and magnified image at center of gravity of the clusters. Purple, yellow, green, and grey sphere correspond to Mo, S, B, and N atom, respectively.